\documentclass[%
hyperref,
reprint,
superscriptaddress,
 amsmath,amssymb,
 aps,
 prb,
 bibnotes,
floatfix,
]{revtex4-2}

\usepackage{graphicx}
\usepackage{dcolumn}
\usepackage{bm}
\usepackage{color}
\usepackage{here}
\usepackage{ulem}
\usepackage{color}

\usepackage[version=4]{mhchem}
\usepackage{xspace}
\newcommand{\tchimax}{$T_{\chi_{\rm max}}$\xspace}
\newcommand{\pc}{$P_{\rm c}$\xspace}
\newcommand{\tc}{$T_{\rm c}$\xspace}
\newcommand{\baxis}{$b$-axis\xspace}

\begin{document}

\preprint{APS/123-QED}

\title{Drastic change in magnetic anisotropy of \ce{UTe2} under pressure revealed by \ce{^125Te}-NMR}

\author{Katsuki Kinjo}

\affiliation{%
{\it Department of Physics, Kyoto University, Kyoto 606-8502, Japan}
}%
\author{Hiroki Fujibayashi}
\affiliation{%
{\it Department of Physics, Kyoto University, Kyoto 606-8502, Japan}
}%
\author{Genki Nakamine}
\affiliation{%
{\it Department of Physics, Kyoto University, Kyoto 606-8502, Japan}
}%
\author{Shunsaku Kitagawa}
\affiliation{%
{\it Department of Physics, Kyoto University, Kyoto 606-8502, Japan}
}%
\author{Kenji Ishida}
\affiliation{%
{\it Department of Physics, Kyoto University, Kyoto 606-8502, Japan}
}%
\author{Yo Tokunaga}%
\affiliation{%
{\it ASRC, Japan Atomic Energy Agency, Tokai, Ibaraki 319-1195, Japan}
}%
\author{Hironori Sakai}
\affiliation{%
{\it ASRC, Japan Atomic Energy Agency, Tokai, Ibaraki 319-1195, Japan}
}%
\author{Shinsaku Kambe}
\affiliation{%
{\it ASRC, Japan Atomic Energy Agency, Tokai, Ibaraki 319-1195, Japan}
}%

\author{Ai Nakamura}
\affiliation{%
{\it IMR, Tohoku University, Oarai, Ibaraki 311-1313, Japan}
}%
\author{Yusei Shimizu}
\affiliation{%
{\it IMR, Tohoku University, Oarai, Ibaraki 311-1313, Japan}
}%
\author{Yoshiya Homma}
\affiliation{%
{\it IMR, Tohoku University, Oarai, Ibaraki 311-1313, Japan}
}%
\author{Dexin Li}
\affiliation{%
{\it IMR, Tohoku University, Oarai, Ibaraki 311-1313, Japan}
}%
\author{Fuminori Honda}
\affiliation{%
{\it IMR, Tohoku University, Oarai, Ibaraki 311-1313, Japan}
}%
\author{Dai Aoki}
\affiliation{%
{\it IMR, Tohoku University, Oarai, Ibaraki 311-1313, Japan}
}%
\affiliation{%
{\it University of Grenoble, CEA, IRIG-PHERIQS, F-38000 Grenoble, France}
}%

\date{\today}

\begin{abstract}
To investigate the normal-state magnetic properties of \ce{UTe2} under pressure, we perform \ce{^125Te} nuclear magnetic resonance (NMR) measurements up to 2~GPa.
Below 1.2~GPa, the \baxis NMR Knight shift shows a broad maximum at the so-called \tchimax on cooling, which is consistent with the magnetization measurement under pressure.
\tchimax decreases with increasing pressure and disappears at the critical pressure \pc = 1.7 GPa, above which superconductivity is destroyed.
This tendency is also observed in the temperature dependence of the nuclear spin-lattice relaxation rate $1/T_1$.
At low pressures, $1/T_1$ shows a conventional Fermi-liquid behavior ($1/T_1T$ = constant) at low temperatures, indicating the formation of the heavy-fermion state.
Above \pc, $1/T_1T$ follows a $1/T$ behavior without any crossover to the heavy-fermion state down to the lowest temperature ($\sim 3$~K).
In addition, the NMR signals disappear below 3~K, due to the influence of the magnetically ordered moments.  
From the pressure dependence of the \tchimax and Knight shift, it was found that the Fermi surface character is abruptly changed at \pc, and that superconductivity is observed only in the heavy-fermion state.

\end{abstract}

\maketitle

The relationship between superconductivity and magnetism has been intensively studied in condensed-matter physics.
This is because magnetism, which was once considered an enemy of superconductivity, is now regarded as a parent state of various unconventional superconductors\cite{MonthouxNature2007}.     
In uranium(U)-based superconductors, such as \ce{UGe2} under pressure ($P$)\cite{SaxenaNature2000}, \ce{URhGe}\cite{AokiNature2001}, and \ce{UCoGe}\cite{HuyPRL2007}, the ground state exhibits the coexistence of ferromagnetism and superconductivity.
The upper critical fields ($H_{\rm{c2}}$) of these superconductors are far beyond the ordinary Pauli-depairing limit, and their superconductivity is enhanced by a magnetic field ($H$) along a specific crystal axis: the superconductivity of UGe$_2$ under pressure is enhanced by $H \parallel a$ (magnetic easy axis)\cite{HuxleyPRB2001}, but that of URhGe\cite{LevyScience2005} and UCoGe\cite{AokiJPSJ2009} is enhanced by $H \parallel b$ (magnetic hard axis).   
Therefore, these are the leading candidates for spin-triplet superconductivity\cite{Aoki2019}.

A new U-based superconductor \ce{UTe2} was discovered and its superconducting (SC) transition temperature ($T_{\rm c}$) is approximately 1.6~K\cite{Ran2019}. 
Although \ce{UTe2} does not exhibit ferromagnetism, \ce{UTe2} is considered to be analogous to ferromagnetic superconductors owing to the following experimental facts: Ising anisotropy in the magnetic susceptibility, $H_{\rm{c2}}$ in all crystal axes exceeding the Pauli-depairing limit, $H$-enhanced superconductivity in $H \parallel b$, and so on\cite{Ran2019,Aoki2019UTe2,aoki2021condmat}.
Recent NMR results in the SC state strongly support the spin-triplet scenario\cite{Nakamine2019, NakamineJPSJ2021, NakaminePRB2021}. 

One of the unique properties of \ce{UTe2} is the broad maximum in the temperature ($T$) dependence of the $b$-axis magnetic susceptibility, the so-called \tchimax \cite{Aoki2019UTe2}.
This is considered to be related to the Kondo coherence temperature\textcolor{black}{, and thus, the \tchimax is regarded as the temperature below which the coherent state in the Kondo lattice (Kondo coherent state) is formed.
Under a high magnetic field along the $b$ axis, \ce{UTe2} shows a sharp metamagnetic transition at 35 T\cite{MiyakeJPSJ2019, RanNatPhys2019}. 
As the relation between \tchimax and the metamagnetic field in \ce{UTe2} follows the well-known empirical relation observed in various heavy-fermion compounds\cite{HiroseJPConf2011}, it is considered that the Kondo coherent state is destroyed at 35~T.
Because superconductivity is also destroyed at the same field, the close relationship between superconductivity and the Kondo coherent state is expected.
Therefore, it is important to investigate the relationship between \tchimax and superconductivity in \ce{UTe2} by tuning other parameters such as hydrostatic or uniaxial pressure.} 

Under hydrostatic pressure, it was reported that its $T_{\rm c}$ strongly increases up to 3~K at $P \sim 1.2 $~GPa, and superconductivity suddenly disappears at critical pressure $P_{\rm c}\sim 1.6$~GPa, above which magnetic anomalies were reported\cite{Braithwaite2019,Ran2020,Thomas2020}.
\textcolor{black}{
The specific heat measurement and the anisotropy of $H_{\rm c2}$ under pressure suggested that the SC character is changed above 0.3~GPa\cite{Braithwaite2019,Ran2020,Thomas2020}. We call the SC state induced by pressure SC2.
Recently, the magnetic susceptibility measurement under pressure revealed the two magnetic anomalies above \pc\cite{AokiJPSJ2021}.
The magnetic-ordered (MO) state is characterized by the sharp change in the magnetic susceptibility, and the weakly-magnetic-ordered (WMO) state is characterized by the broad maximum of the susceptibility for $H \parallel a$ and $c$.
The magnetic short-range correlations are considered to develop in the WMO state\cite{AokiJPSJ2021}.
The magnetic characters of these two states are still under investigation.
}
Although the pressure-induced phases in \ce{UTe2} have attracted a lot of attention, few experimental results have been reported \cite{Braithwaite2019,Thomas2020,Li2021} owing to the difficulty of experiments.
\textcolor{black}{NMR under pressure is a microscopic measurement that can investigate magnetic properties via the hyperfine fields at the NMR-nuclear site without background subtraction from a pressure cell.}

Here, we report the results of the NMR Knight shift ($K$) and the nuclear spin--lattice relaxation rate $1/T_1$ in $H \parallel b$- and $c$-axes under pressure.
Below $P_{\rm c}$, \tchimax decreases with increasing pressure, which is consistent with previous magnetization measurements\cite{Knebel2020}.
Above $P_{\rm c}$, the $b$-axis $K$ ($K_b$) is strongly enhanced and exhibits Curie--Weiss behavior in the $T$ dependence. 
In contrast, the $c$-axis $K$ ($K_c$) is almost pressure-independent, but shows a maximum in $T$ dependence above $P_{\rm c}$.
The $P$ dependence of the Knight shift reveals that the low-temperature magnetic anisotropy is very sensitive to the applied pressure. 
\textcolor{black}{In addition, the NMR signals disappear below 3~K under pressure above \pc, indicating a magnetic ground state.}
This is consistent with previous results\cite{Braithwaite2019,Ran2020,Thomas2020}.

Single-crystal \ce{UTe2} was grown using the chemical transport method with I as a transport agent with 99.9\% enriched $^{125}$Te and natural U as the starting elements.
\ce{^125Te}-NMR measurements were performed on a single crystal of $2 \times 1 \times 1$ mm$^3$ size. 
The \ce{^125Te} nucleus has a spin of 1/2 with a gyromagnetic ratio $^{125}\gamma /2\pi$ = 13.454~MHz/T. 
The \ce{^125Te}-NMR spectrum was obtained using the Fourier transform of a spin-echo signal observed after the spin-echo radio frequency pulse sequence. 
We used a split SC magnet generating a horizontal magnetic field and a single-axis rotator with the $a$ axis as the rotation axis to apply $H$ exactly parallel to the $b$ and $c$ axes, as shown in the inset of Fig.~1(c).
1/$T_1$ of \ce{^125Te} was determined by fitting the time dependence of the spin-echo intensity $M(t)$ at $t$ after saturation of the nuclear magnetization $M_0$. 
The fitting function is a single exponential function for $I$ = 1/2.
Hydrostatic pressure was applied with a piston-cylinder-type pressure cell\cite{Uwatoko2002}, and Daphne 7373 was used as the pressure medium.
The applied pressures were estimated from the \tc of Pb from the resistivity measurements \cite{Bireckoven1988}.
Because there are two inequivalent Te sites in UTe$_2$, as shown in Fig.~1(a), two NMR peaks were observed when $H$ was applied to the $b$-axis, as shown in Fig.~1(c). 
Although we measured both sites, we focused on the Te(1) peak with a smaller Knight shift in $H \parallel b$, as the NMR results are essentially the same at the two sites.

\begin{figure}[h]
\includegraphics[width = \linewidth]{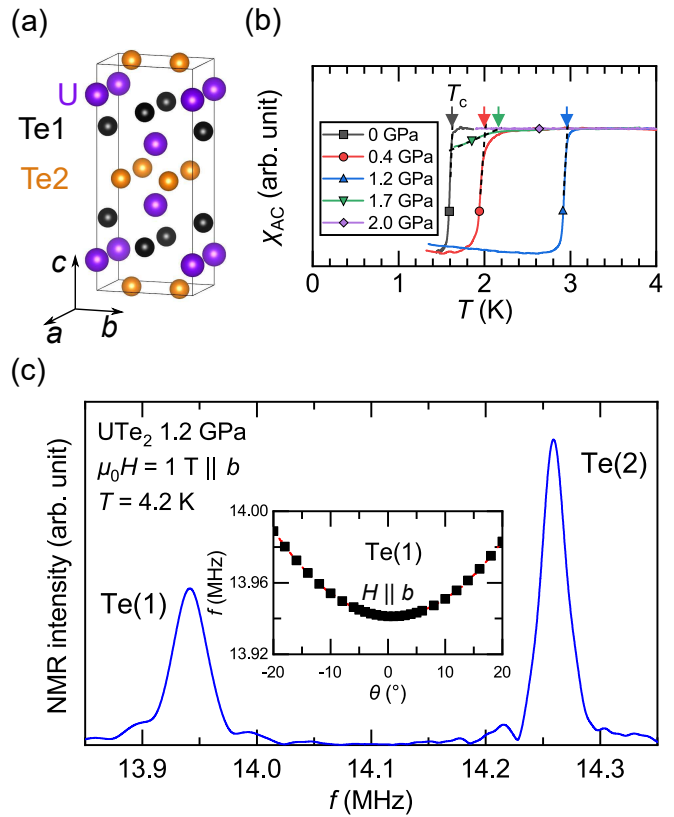}
\caption{(Color online) (a) Crystal structure of \ce{UTe2} made by VESTA \cite{VESTA}.
(b) $T$ dependence of magnetic AC susceptibility $\chi_{\rm AC}$ at several pressures. The measurement was performed using an NMR tank circuit on cooling.
$T_{\rm c}$'s are shown by the arrows.
(c) Typical \ce{^125Te}-NMR spectrum measured in the field along the $b$ axis. (inset) Angular dependence of the resonance frequency of the NMR signal at the Te(1) site around $H \parallel b$. \textcolor{black}{$\theta = 0$ represents $H \parallel b$.} The red dashed line represents the fitting with the calculation.}
\label{fig1_new}
\end{figure}

Figure~\ref{fig1_new}(b) shows the $T$ dependence of the AC magnetic susceptibility ($\chi _{\rm AC}$) under various pressures, and $T_{\rm c}$'s are shown by arrows.
$T_{\rm c}$ first increases and reaches a maximum at approximately 1.2 GPa with increasing pressure, and then superconductivity suddenly disappears above $P_c$.
The $P$ dependence of the present $T_{\rm c}$ is in good agreement with previous results\cite{Braithwaite2019,Ran2020}, and the $P_{\rm c}$ of this sample was determined to be $\sim$ 1.7 GPa.
This is because the sharp transition ascribed to the bulk Meissner signal was not observed, although a relatively small decrease in $\chi_{\rm AC}$ was observed at 1.7 GPa. No Meissner signal was observed at $P = 2.0$ GPa.

\begin{figure}[tp]
\includegraphics[width = \linewidth]{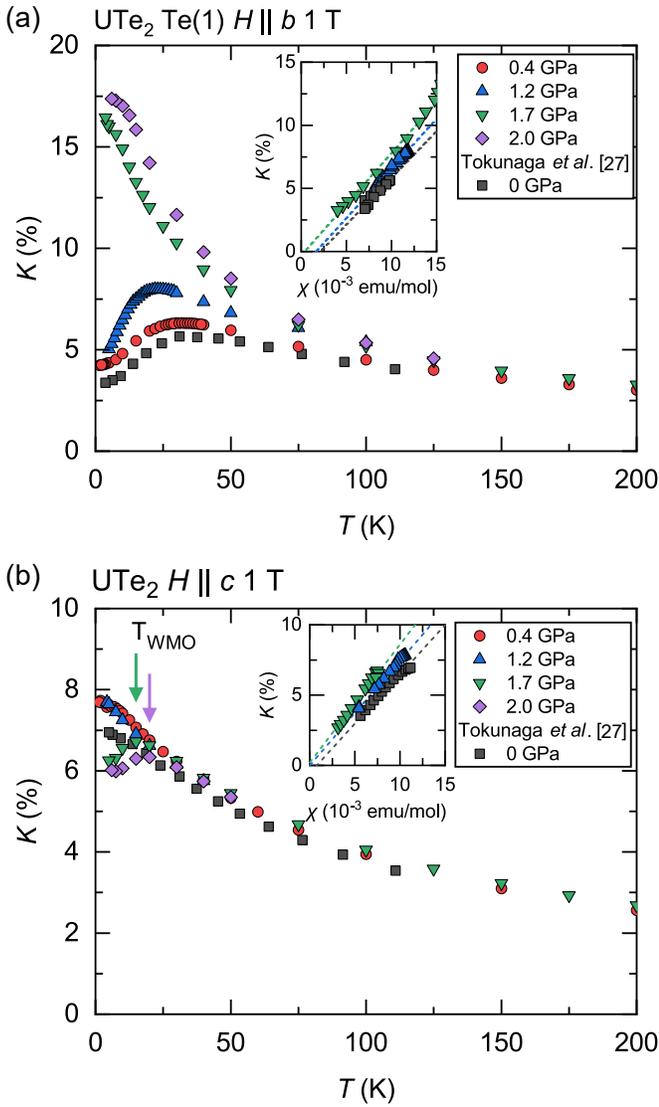}
\caption{(Color online)
Temperature dependence of the NMR Knight shift $K$ for (a) $H \parallel b$ and (b) $H \parallel c$.
Below 1.7 GPa, i.e., \pc, $K_b$ shows a broad maximum in $T$ dependence.
In contrast, above \pc, $K$ increases with decreasing temperature down to the lowest temperature.
(inset) $K$-$\chi$ plots for 0 GPa, 1.2 GPa, and 1.7 GPa for (a) $H \parallel b$ and (b) $H \parallel c$. The value of $\chi$ under pressure was reported in ref. \cite{Li2021}.
The dashed lines in the figure is linear-fitting line.
}
\label{fig2}
\end{figure}
\begin{table}[]
\vspace{5mm}
\centering
\caption{\label{table:hfconstant}The hyperfine coupling constant ($A_{\rm hf}$) of \ce{UTe2} at several pressures. $A_{\rm hf}^{1[2]}$ is the hyperfine coupling constant of Te(1) [Te(2)]. $A_{\rm hf}$ of Te(1) and Te(2) for $H \parallel c$ has almost same value and are indistinguishable.}
\begin{tabular}{ccc}
\multicolumn{3}{c}{$H \parallel b$} \\[0.5mm]
\hline \hline 
$P$ (GPa) & $A_{\rm hf}^{1}$ (T/$\mu_{\rm B}$) & $A_{\rm hf}^{2}$ (T/$\mu_{\rm B}$) \\
\hline
0  & $4.1 \pm 0.3$ & $5.6 \pm 0.4$\\
1.2  & $4.3 \pm 0.3$ & $5.6 \pm 0.4$\\
1.7  & $4.4 \pm 0.5$& $5.7 \pm 0.6$ \\
\hline \hline
\end{tabular}
\begin{tabular}{cc}
\multicolumn{2}{c}{$H \parallel c$} \\[0.5mm]
\hline \hline $P$ (GPa) & $A_{\rm hf}$ (T/$\mu_{\rm B}$) \\
\hline
0 & $3.9 \pm 0.2$ \\
1.2 & $4.0 \pm 0.2$  \\
1.7 & $4.4 \pm 0.3$  \\
\hline \hline
\end{tabular}
\end{table}

We found a drastic change in the $T$ dependence of $K_b$ around \pc.
Figures \ref{fig2}(a) and \ref{fig2}(b) show the $T$ dependencies of $K_b$ and $K_c$, respectively.
 \textcolor{black}{Usually, the NMR Knight shift and the susceptibility $\chi$ follow the relation,
$K = A_{\rm hf} \chi + b \chi_0$, where $A_{\rm hf}$ is the hyperfine coupling constant and $b\chi_0$ term is the temperature-independent background contribution.
In this case, the background contribution mainly comes from the bulk-susceptibility of the pressure cell.
In fact, good linear relations between the two quantities were confirmed down to the lowest temperature beyond a broad maximum in the previous ambient-pressure susceptibility\cite{Tokunaga2019}.}
We performed the $K-\chi$ plot in $P = $~1.2 and 1.7~GPa, which is shown in the insets of Fig.~2.
\textcolor{black}{The hyperfine coupling constant estimated from the slope of the $K$-$\chi$ plot is summarized in Table I, and is almost $P$-independent}; the $P$ dependence of the NMR physical quantities is reasonably ascribed to that of the magnetic-susceptibility properties. 
With increasing pressure, the \baxis \tchimax gradually shifts to a low temperature, which is consistent with the results of the AC and DC magnetic susceptibility measurements\cite{Knebel2020}.
Above \pc, $K_b$ does not show the broad maximum and diverges to the lowest temperature.


\begin{figure}[H]
\includegraphics[width = \linewidth]{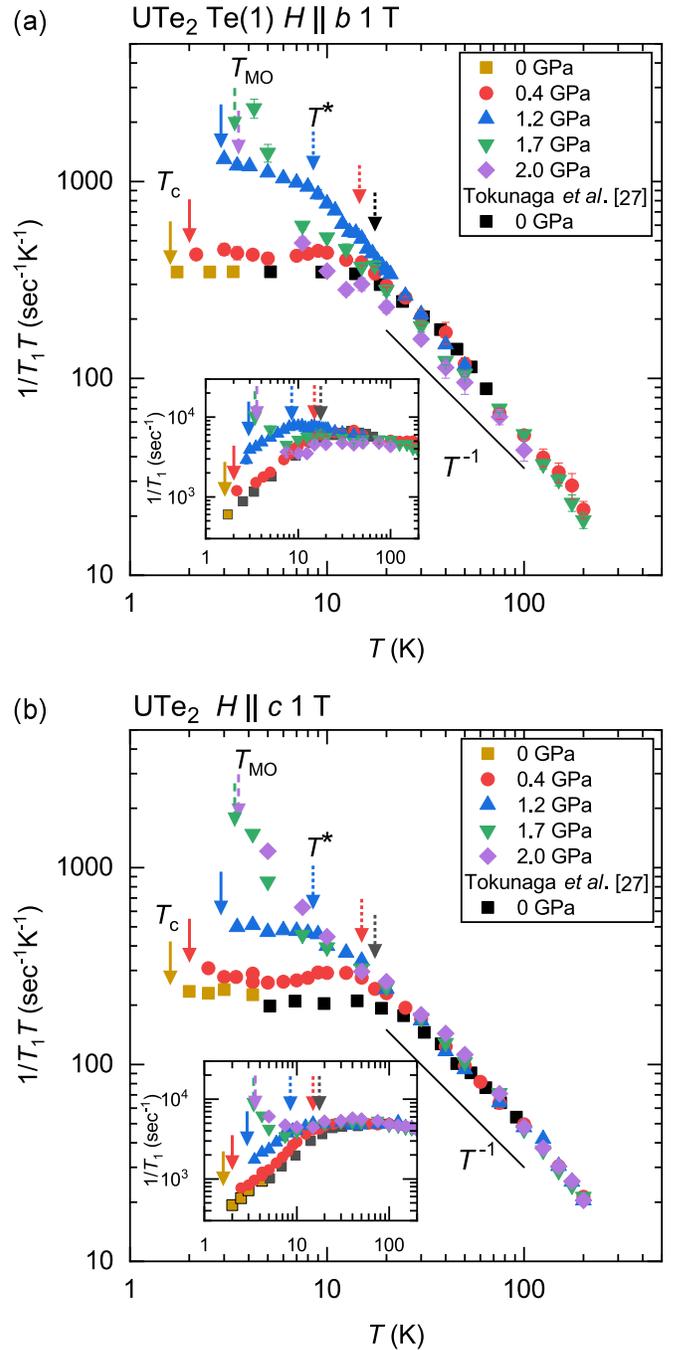}
\caption{(Color online)
$T$ dependence of the nuclear spin--lattice relaxation rate divided by $T$ ($1/T_1T$), which is proportional to the $q$-averaged dynamical spin susceptibility in (a) $H \parallel b$ and (b) $H \parallel c$.
The insets are the $T$ dependence of $1/T_1$ for $H \parallel b$ and $H \parallel c$.
$T_{\rm c}$, $T_{\rm MO}$, and $T^*$ are shown by solid, dashed, and dotted arrows, respectively.
}
\end{figure}

In contrast, the absolute value of $K_c$ is almost the same at all pressures, and thus the magnetic hardest axis at low temperatures changes from the $b$ to $c$ axis above $P_c$.
In addition, an appreciable peak appears in the temperature dependence above \pc.
It was reported that, above \pc (1.7 GPa), the magnetic susceptibility along the $c$-axis ($\chi_c$) shows a broad maximum at $T_{\rm WMO} \sim$ 10 K, where a magnetic short-range order develops\cite{Li2021}.
Thus, the peak in $K_c$ corresponds to the peak in $\chi_c$ at $T_{\rm WMO}$, and this gradually shifts to a higher temperature with increasing pressure.
These results indicate that the origin of the peak in $K_c$ is different from that in $K_b$ at \tchimax.

\textcolor{black}{Figures 3(a) and 3(b) show the $T$ dependence of $1/T_1T$, which is proportional to the $q$-averaged dynamical spin susceptibility in (a) $H \parallel b$ and (b) $H \parallel c$. 
In contrast to the Knight shift, the $T$ and $P$ dependences of $1/T_1T$ are almost the same in both directions, although $1/T_1T$ behavior below $T^*$ is slightly different at 1.2 GPa.
This is because $1/T_1T$ probes magnetic fluctuations perpendicular to the applied magnetic field and both $1/T_1T$'s are governed by fluctuations arising from the magnetic easy-axis ($a$-axis) component\cite{Tokunaga2019}.    
At ambient pressure, $1/T_1T$ exhibits the typical heavy-fermion behavior as observed in UPt$_3$\cite{KohoriJPSJ1988}. 
$1/T_1T$ is proportional to $1/T$ in the high-temperature region owing to the localized 5$f$-electron character.
On cooling, $1/T_1T$ deviates from the $1/T$ behavior owing to the hybridization between the localized 5$f$-electron and conducting U-6$d$ and Te-5$p$ electrons\cite{aoki2021condmat}.
The U- 5$f$ electrons show an itinerant character below the Kondo coherent temperature characterized by \tchimax, and $1/T_1T$ becomes almost constant below $T^*$, which is a characteristic behavior of the heavy-fermion state.
When pressure is applied, the \tchimax decreases and the Korringa value (the constant value of $1/T_1T$ at low $T$) increases, indicating that the bandwidth becomes narrower and the electron correlation becomes stronger.
In contrast, $1/T_1T$ continues to increase down to the lowest $T$ above \pc, analogous to the $K_b$, and the heavy-fermion state was no more observed down to 3~K.
It indicates that the heavy-fermion state and the Kondo coherent state were destroyed above \pc.}

From the present results, as well as those previously reported\cite{Braithwaite2019, Ran2020, Knebel2020, Li2021}, we developed the 
$P$--$T$ phase diagram shown in Fig.~4(b).
This phase diagram shows that the Kondo coherent state was destroyed above \pc, where the U 5-$f$ state remained in the localized state down to 3 K.
It was clarified that superconductivity is observed only in the Kondo coherent state, where the U 5-$f$ state becomes itinerant.
We point out the similarity between the $P$ dependence and $H \parallel b$ phase diagrams because the superconductivity is enhanced by the application of $P$ or $H \parallel b$ but is abruptly suppressed by the destruction of the Kondo coherent state in both cases.
\textcolor{black}{Above \pc or metamagnetic field, magnetic properties are not well-understood so far.
The gradual increase in $K_b$ at low $T$ by applying a small pressure [Fig.~4(a)] is the same tendency as the $H$ dependence of $K_b$, in which $K_b$ just above $T_c$ slightly increases with increasing $H$ up to 15 T\cite{NakamineJPSJ2021}.}
It is considered that the gradual increase in $K_b$ and gradual decrease in \tchimax with pressure lower the metamagnetic fields in $H \parallel b$, as was actually observed in the high-field magnetization measurement under pressure\cite{Knafo2021}.
In addition, $K_b$ at low $T$ discontinuously increases at \pc, \textcolor{black}{suggesting that the evolution of the electronic state at \pc looks like a first-order transition. }

Above \pc, the NMR signals of Te(1) and Te(2) disappeared below $T_{\rm MO}$ assigned by the resistivity and magnetization measurements\cite{Li2021,AokiJPSJ2021} because of the influence of the magnetic ordered moments. 
Recently, inelastic neutron scattering measurements reported that \textcolor{black}{the magnetic fluctuations along the $a$ axis with ${\bm Q}_{\rm INS} = (0,0.57,0)$ are roughly proportional to $1/T_1T$ at ambient pressure\cite{Duan2020,KnafoINS2021}}.
If this relation is still preserved even above \pc, $1/T_1T$ diverges toward $T_{\rm MO}$, indicating that the MO state is the AFM ordered state with incommensurate ${\bm Q}_{\rm INS}$.
In such a case, the NMR spectrum is smeared out due to distributed internal fields at the Te sites.

\begin{figure}[H]
\includegraphics[width = \linewidth]{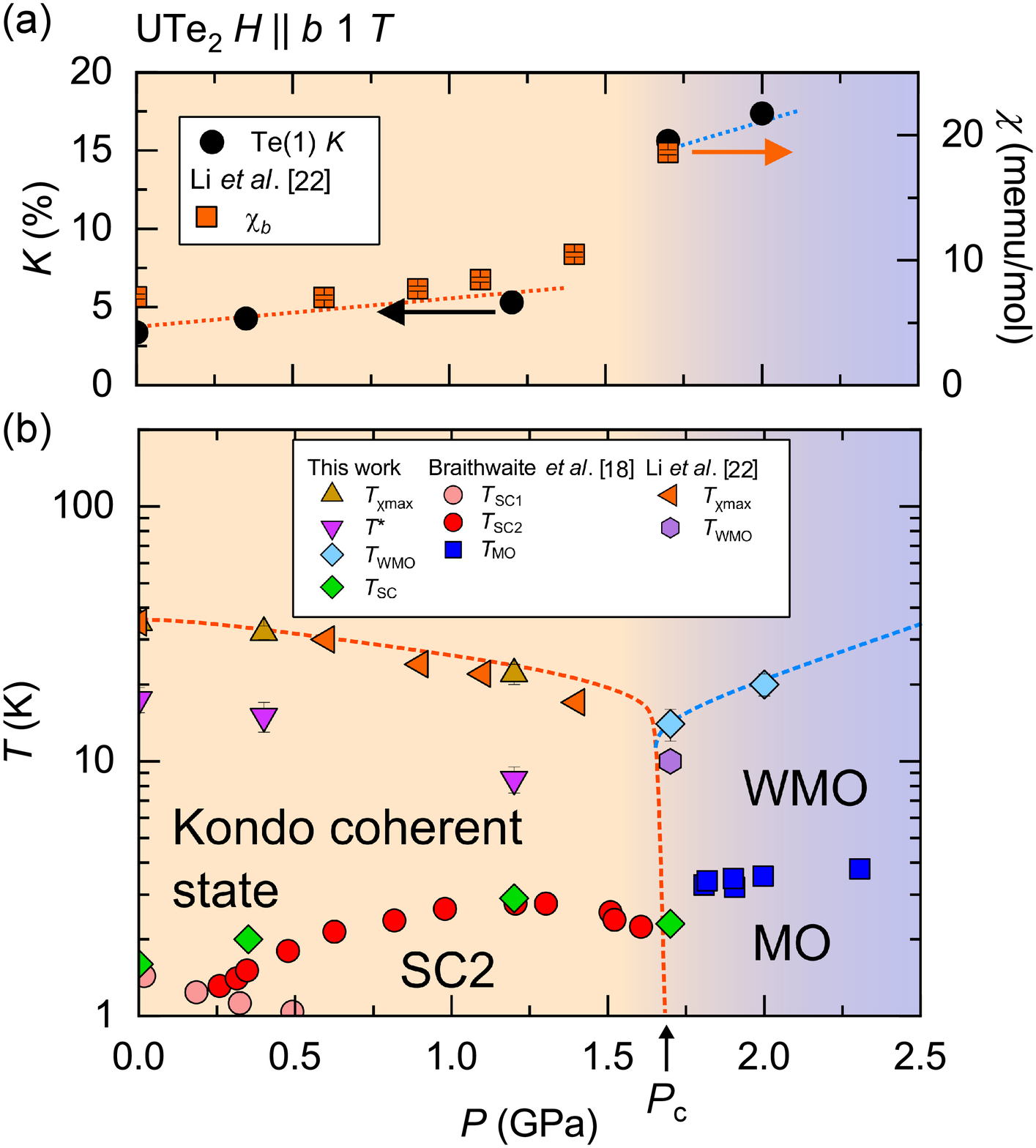}
\caption{(Color online)
(a) $P$ dependence of the NMR Knight shift at 6~K. The value of the Knight shift suddenly increases around 1.7~GPa with increasing pressure.
(b) $P$--$T$ phase diagram of \ce{UTe2}.
$T_{\rm SC}$ is the superconducting critical temperature.
\tchimax is the temperature peak position in the $b$-axis Knight shift and $T^*$ is the crossover temperature, below which $1/T_1T$ is almost constant. \tchimax and $T^*$ decrease with increasing pressure and suddenly disappear above \pc.
$T_{\rm WMO}$ is the peak temperature in the $c$-axis Knight shift and $\chi$.}
\end{figure}

In conclusion, from the $P$ dependence of the NMR results in the $H \parallel b$- and $c$-axes, it was found that $K_b$ and $1/T_1T$ at low $T$ discontinuously increase at \pc, suggesting an abrupt change in the electronic state and the breakdown of the Kondo coherent state above \pc.
We observed the similarity between the pressure and $H \parallel b$ phase diagrams.  
It was revealed that superconductivity can be realized only in the Kondo coherent state, indicating that the itinerant 5-$f$ nature is crucial for superconductivity in UTe$_2$.
Despite the similarity of the electronic states in the normal state and the SC properties between UTe$_2$ and other U-based FM superconductors, the relationship between the magnetic phase and the superconductivity is quite different. 
Our study clarifies the richness of the spin-triplet superconductors even in the U-based superconductors.



The authors would like to thank M. Manago, J. Ishizuka, Y. Yanase, Y. Maeno, S. Yonezawa, and J-P. Brison, G., Knebel, and J. Flouquet for valuable discussions, and Editage (www.editage.com) for English language editing. This work was supported by the Kyoto University LTM Center, Grants-in-Aid for Scientific Research (Grant Nos. JP15H05745, JP17K14339, JP19K03726, JP16KK0106, JP19K14657, JP19H04696, JP19H00646, JP20H00130, and JP20KK0061), and a Grant-in-Aid for JSPS Research Fellows (Grant No. JP20J11939) from JSPS.

\bibliographystyle{apsrev4-2}

\end{document}